\documentclass[conference, a4paper]{IEEEtran}
\IEEEoverridecommandlockouts
\usepackage{cite}
\usepackage{amsmath,amssymb,amsfonts}
\usepackage{algorithmic}
\usepackage{graphicx}
\usepackage{textcomp}
\usepackage{xcolor}
\usepackage{bm}
\def\BibTeX{{\rm B\kern-.05em{\sc i\kern-.025em b}\kern-.08em
    T\kern-.1667em\lower.7ex\hbox{E}\kern-.125emX}}

\newtheorem{lemm}{Lemma}
\newtheorem{rema}{Remark}

\begin{document}

\title{Covariance Evolution for Spatially ``Mt. Fuji'' Coupled LDPC Codes
\thanks{This work was supported by JSPS KAKENHI Grant Numbers JP16K00195, JP16K00417, JP17K00316, JP17K06446, JP18K11585}
}

\author{\IEEEauthorblockN{Yuta Nakahara}
\IEEEauthorblockA{\textit{Center for Data Science} \\
\textit{Waseda University}\\
	27 Waseda-cho, Shinjuku-ku, Tokyo, 162-0042 Japan\\
	E-mail: yuta.nakahara@aoni.waseda.jp}
\and
\IEEEauthorblockN{Toshiyasu Matsushima}
\IEEEauthorblockA{\textit{Department of Pure and Applied Mathematics} \\
\textit{Waseda University}\\
	3-4-1 Okubo, Shinjuku-ku, Tokyo, 169-8555 Japan\\
	E-mail: toshimat@waseda.jp}
}

\maketitle

\begin{abstract}
A spatially ``Mt.\ Fuji'' coupled low-density parity check (LDPC) ensemble is a modified version of the original spatially coupled (SC) LDPC ensemble. Its desirable properties are first observed in experimentally. The decoding error probability in the error floor region over the binary erasure channel (BEC) is theoretically analyzed later. In this paper, as the last piece of the theoretical analysis over the BEC, we analyze the decoding error probability in the waterfall region by modifying the covariance evolution which has been used to analyze the original SC-LDPC ensemble.
\end{abstract}

\begin{IEEEkeywords}
spatially coupled codes, covariance evolution, finite-length code performance
\end{IEEEkeywords}

\section{Introduction}
A spatially coupled (SC) low-density parity check (LDPC) ensemble\cite{kudekar} is constructed as a set of random bipartite graphs like a chain of $(2L+1)$ block LDPC codes whose code lengths are $M$. Then, the code length of the SC-LDPC ensemble is $N = (2L+1)M$. If $M$ and $L$ are sufficiently large, the SC-LDPC ensemble has many desirable properties. In particular, the belief propagation (BP) threshold $\epsilon^{\mathrm{BP}}$ of the SC-LDPC ensemble coincides with the maximum a posteriori (MAP) threshold of the underlying block LDPC ensemble for sufficiently large $M$. Moreover, the design rate of the SC-LDPC ensemble converges to the design rate of the underlying block LDPC ensemble for sufficiently large $L$ with $O(L^{-1})$. Note that the design rate is independent of $M$. 

However, some problems occur when the code length $N = (2L+1)M$ is fixed to a finite value. In order to increase the design rate, we have to increase $L$ and decrease $M$. If $L$ is too large, the average number of iterations of BP decoding increases in the waterfall region. If $M$ is too small, the decoding error probability increases in the error floor region.\cite{olmos_isit}

For this problems, a generalized SC-LDPC ensemble has been proposed, which is called spatially ``Mt.\ Fuji'' coupled (SFC) LDPC ensemble\cite{SFC}. In the SFC-LDPC ensemble, code lengths of the underlying LDPC codes are different from each other. As the position of the underlying code gets close to the middle of the chain, its code length increases exponentially. The increasing rate is expressed by a parameter $\alpha \geq 1$. Therefore, the design rate of the SFC-LDPC ensemble converges to the design rate of the underlying LDPC ensemble with $O(\alpha^{-L})$ as $L \to \infty$. In the rest of this section, we assume that the design rate and the code length of the SFC-LDPC ensemble and those of the SC-LDPC ensemble are equal to each other and $\alpha > 1$. Then, $L$ of the SFC-LDPC ensemble becomes smaller and $M$ of the SFC-LDPC ensemble becomes larger than those of the SC-LDPC ensembles. 

Studies of the decoding performance of the SFC-LDPC ensemble has been started from the binary erasure channel (BEC).
In the error floor region, the decoding error probability of the SFC-LDPC ensemble is lower than that of the SC-LDPC ensemble. That is first observed experimentally in \cite{SFC}. Later, that is theoretically explained by a weight distribution analysis for the SFC-LDPC ensemble in \cite{SFC_weight_distribution}.

The average number of iterations of the SFC-LDPC ensemble in the waterfall region is lower than that of the SC-LDPC ensemble. That is theoretically expected by an observation of the ``decoding wave'' given by the density evolution and confirmed by numerical experiments in \cite{SFC}.

The decoding error probability of the SFC-LDPC ensemble in the waterfall region is lower than or almost equal to that of the SC-LDPC ensemble if $\alpha$ is appropriately tuned. If $\alpha$ is too large, the decoding error probability becomes larger. This phenomenon is observed experimentally in \cite{SFC}. It is guessed in \cite{SFC} as an effect of decrease of the BP threshold $\epsilon^{\mathrm{BP}}$, which is regarded as an asymptotic indicator of the waterfall performance. However, the decoding error probability under a finite code length has not been theoretically analyzed yet.

For the original SC-LDPC ensemble, the finite-length decoding error probability in the waterfall region over the BEC is analyzed by combining two systems of differential equations called expected graph evolution (EGE) and covariance evolution (CE). The EGE for the SFC-LDPC ensemble has been proposed in \cite{SFC_EGE}. In this paper, we derive the CE for the SFC-LDPC ensemble and combine them. Then, we explain the above phenomenon more theoretically and directly than in previous studies.

\section{Preliminary}
At first, we describe some notations. Let $\mathbb{N}$, $\mathbb{Z}$, and $\mathbb{Q}$ denote the set of natural numbers, integers, and rational numbers, respectively. For any integers $i$ and $j$, $(i < j)$, let $[i, j]$ denote the set of $\{ i, i+1, \dots , j\}$. Let $\lceil \cdot \rceil$ denote the ceiling function.

\subsection{$(d_v, d_c, L, \alpha)$ spatially ``Mt.\ Fuji'' coupled LDPC ensemble}
In this section, we describe the spatial coupling of LDPC codes with increasing code length. The constructed ensemble is called a spatially ``Mt.\ Fuji'' coupled (SFC) LDPC ensemble. In the SFC-LDPC ensemble, the code length of LDPC code at position $i \in [-L, L]$ is $\lceil \alpha^{L-|i|} M \rceil$ ($\alpha \in \mathbb{Q}$, $\alpha \geq 1$, $M \in \mathbb{N}$). This is in contrast to the usual SC-LDPC ensemble, where the code length of every LDPC code is $M$. Although we can define an ensemble like \cite{kudekar} with smoothing parameter $w$, we describe only the definition of an ensemble like \cite{olmos}, which is suitable for finite-length analysis.

A $(d_v, d_c, L, \alpha)$ SFC-LDPC ensemble is defined as a set of random bipartite graphs which are constructed by the following 4 steps.

\begin{enumerate}
\item Set variable nodes\mbox{}\\
At position $i \in [-L, L]$, $L \in \mathbb{N}$, $\lceil \alpha^{L-|i|} M \rceil$ variable nodes of degree $d_v \in \mathbb{N}$ are set. At position $i \in [-L-d_v+1,-L-1] \cup [L+1, L+d_v-1]$, $\lceil \alpha^{L-|i|} M \rceil$ dummy nodes of degree $d_v$ are set. The dummy nodes are shortened at the last step. 

\item Extend edges deterministically\mbox{}\\
The $j$th $(j \in [0, d_v-1])$ edge of the variable (or dummy) node at position $i \in [-L-d_v+1, L+d_v-1]$ is extended to the position $i+j$. Therefore, each variable (or dummy) node extends just one edge to each of the next $d_v$ positions deterministically.

\item Set check nodes\mbox{}\\
Because of the above steps, $\sum_{j=0}^{d_v-1}\lceil \alpha^{L-|i-j|} M \rceil$ edges come from variable nodes at positions $i, i-1, \dots , i-d_v+1$ to the position $i$ of the check node side. Then, we set $\left\lceil\frac{1}{d_c}\sum_{j=0}^{d_v-1}\lceil \alpha^{L-|i-j|} M\rceil \right\rceil$ check nodes at position $i \in [-L, L+d_v-1]\, (d_c \in \mathbb{N})$. In order to equalize the number of edges, only one check node has degree 
\begin{align}
r_i = &\sum_{j=0}^{d_v-1}\lceil \alpha^{L-|i-j|} M \rceil \nonumber \\
&\quad - d_c\left(\left\lceil \frac{1}{d_c}\sum_{j=0}^{d_v-1}\lceil \alpha^{L-|i-j|} M \rceil \right\rceil-1 \right)
\end{align}
and the others have degree $d_c$.

\item Connect edges probabilistically\mbox{}\\
At each position, the edges of the check nodes are connected to the variable or dummy nodes according to a random permutation of $\sum_{j=0}^{d_v-1}\lceil \alpha^{L-|i-j|} M\rceil$ letters. Finally, dummy nodes are shortened.
\end{enumerate}

\begin{rema}\label{alphaml}
Under a fixed code length and a design rate, as $\alpha$ increases, $M$ increases, and $L$ decreases because of the above definition\cite{SFC}.
\end{rema}

\subsection{Probabilistic properties of the ensemble}
If $M$ is sufficiently large and $\alpha^{L-|i|} M$ and both $\frac{1}{d_c}\sum_{j=0}^{d_v-1} \alpha^{L-|i-j|} M$ are natural number, the following lemma holds, where sampling without replacement of edges are approximated by sampling with replacement.

\begin{lemm}[Probabilistic property of $(d_v, d_c, L, \alpha)$ SFC-LDPC ensemble]
\begin{enumerate}
\item The $j$th ($j \in [0,d_v-1]$) edge of a variable node at position $i \in [-L, L]$ is connected to a check node at position $i+j$ with probability $1$.

\item An edge of a check node at position $i \in [-L, L+d_v-1]$ is connected to a variable or dummy node at position $i-j$, $j \in [0, d_v-1]$ with probability $\alpha^{L-|i-j|} / \sum_{k=0}^{d_v-1} \alpha^{L-|i-k|}$.

\item An edge of a check node at position $i \in [-L, L+d_v-1]$ is connected to a variable (not dummy) node in the range of positions $[i-d_v+1, i]$ with probability $s_{i, \alpha} = \frac{\sum_{k=\max \{ -L, i-d_v+1\}}^{\min \{ L, i \}}\alpha^{L-|k|}}{\sum_{j=0}^{d_v-1}\alpha^{L-|i-j|}}$.

\item At least one edge of a check node at position $i \in [-L, L+d_v-1]$ is connected to a variable (not dummy) node in the range of positions $[i-d_v+1, i]$ with probability $(1- ( 1 - s_{i, \alpha} )^{d_c})$.

\item A check node at position $i$ has a degree $m$ with probability
\begin{align}
&\rho_{m,i, \alpha} = \nonumber \\
&\begin{cases}
1, \quad i \! \in \! [-\!L\!+\!d_v\!-\!1, L], m=d_c, \\
0, \quad i \! \in \! [-\!L\!+\!d_v\!-\!1, L], m < d_c, \\
\binom{d_c}{m} \left( s_{i, \alpha} \right)^{m} \left( 1 - s_{i, \alpha} \right)^{d_c-m},\\
\qquad i \! \in \! [-\!L, \!-\!L\!+\!d_v\!-\!2] \! \cup \! [L\!+\!1, L\!+\!d_v\!-\!1].
\end{cases}\label{lem1-5}
\end{align}
\end{enumerate}
\end{lemm}

\section{Covariance evolution for the SFC-LDPC ensemble}\label{sec3}

In this section, we describe the CE for the SFC-LDPC ensemble in order to analyze the decoding error probability in the waterfall region. In the following, we assume that codewords are transmitted through the BEC with channel erasure probability $\epsilon$ (BEC($\epsilon$)). In addition, the peeling decoder\cite{luby} is assumed in the analysis. It has the same decoding error probability as the BP decoder in a sufficiently large number of iterations. Let $t$ denote the iteration number of the peeling decoder. Let $V_u(t)$ denote the number of variable nodes at position $u \in [-L, L+d_v-1]$ in the residual graph. Let $R_{j,u} (t)$ denote the number of edges connected to the check nodes of degree $j \in [1, d_c]$ at the position $u \in [-L, L+d_v-1]$ in the residual graph. And their normalized versions are defined by $\tau = t/M$, $v_u(\tau) = V_u(\tau M)/M$, and $r_{j,u}(\tau) = R_{j,u}(\tau M)/M$. 

As $M \to \infty$, the expected behavior $\hat{v}_u(\tau) = \mathbb{E}[v_u(\tau)]$ and $\hat{r}_{j,u}(\tau) = \mathbb{E}[r_{j,u}(\tau)]$ of the peeling decoder for the SC-LDPC ensemble over the $\mathrm{BEC}(\epsilon)$ is known to satisfy a system of differential equations called expected graph evolution (EGE)\cite{olmos}, where the expectation is taken over the ensemble, channel outputs, and the random choice of a degree 1 check node made by the peeling decoder. In addition, let $\delta^{i,j}_{z,x}(\tau) = \mathrm{CoVar} [r_{i,j}(\tau), r_{z,x}(\tau)] M$. $\delta^{i,j}_{z,x}(\tau)$ is known to satisfy a system of differential equations called covariance evolution (CE)\cite{olmos} as $M \to \infty$. Moreover, $r_{j, u} (\tau)$ is Gaussian distributed with mean $\hat{r}_{j,u}(\tau)$ and variance $\delta_{j,u}^{j,u}(\tau)/M$ for a sufficiently large $M$\cite{olmos}.

Since the decoding rule for the SFC-LDPC ensemble is the same as that for the SC-LDPC ensemble, the difference between the SFC-LDPC ensemble and the SC-LDPC ensemble appears in the initial conditions of the EGE and the CE. The initial condition of the EGE for the SFC-LDPC ensemble has been proposed in \cite{SFC_EGE}. We reproduce it in the Appendix \ref{A}. In this paper, we derive the initial conditions of the CE for the SFC-LDPC ensemble. In order to confirm that our modification\footnote{the $(l-c)\frac{ab}{M}$ at formula (97) in \cite{olmos} is probably mistake of $\frac{ab}{(l-c)M}$. Our modification is based on the latter term.} of the initial conditions is a natural generalization of those for the SC-LDPC ensemble, we describe the initial condition only for the most difficult case where $u, x \in [-L, L+d_v-1]$, $j, z \in [1, d_c]$, $u < x$, $|u-x| < d_v$. The other initial conditions and their derivations are in the Appendix \ref{B}

\begin{align}
&\delta_{z,x}^{j,u}(0) \nonumber \\
&= \mathrm{CoVar}[R_{j,u}(0), R_{z,x}(0)] / M  \nonumber \\
& = jz \left( \sum_{k= \max \{-L, x-d_v+1\}}^{\min \{ L, u\}} \alpha^{L-|k|} \right) \nonumber \\
&\times \Bigl( P(d_u \!=\! j, d_x \!=\! z | \mathrm{share}) \!-\! P(d_u \!=\! j, d_x \!=\! z | \mathrm{no \ share}) \Bigr).
\end{align}
where
\begin{align}
P(&d_u = j, d_x = z | \mathrm{share}) =\nonumber \\
& \epsilon \left[ \left( \sum_{m=j}^{d_c} \rho'_{m,u,\alpha} \binom{m-1}{j-1} \epsilon^{j-1} (1-\epsilon)^{m-j} \right)\right. \nonumber \\
& \left. \times \left( \sum_{m = z}^{d_c} \rho'_{m, x,\alpha} \binom{m-1}{z-1} \epsilon^{z-1} (1-\epsilon)^{m-z} \right) \right] \nonumber \\
& + (1- \epsilon) \left[ \left( \sum_{m = j+1}^{d_c} \rho'_{m,u,\alpha} \binom{m-1}{j} \epsilon^j (1- \epsilon)^{m-j-1} \right)\right. \nonumber \\
& \left. \times \left( \sum_{m = z+1}^{d_c} \rho'_{m,x,\alpha} \binom{m-1}{z} \epsilon^z (1-\epsilon)^{m-z-1} \right) \right],
\end{align}
\begin{align}
&\rho'_{m,u, \alpha} =\nonumber \\
&\begin{cases}
1, \quad u \in [-L+d_v-1, L], m=d_c,\\
0, \quad u \in [-L+d_v-1, L], m < d_c,\\
\binom{d_c-1}{m-1} \left( s_{u,\alpha} \right)^{m-1} \left( 1 - s_{u, \alpha} \right)^{d_c-m},\\
\quad u \in [-L, -L+d_v-2] \cup [L+1, L+d_v-1].
\end{cases}
\end{align}
\begin{align}
&P(d_u = j, d_x = z | \mathrm{no \ share}) \nonumber \\
&=\left( \sum_{m = j}^{d_c} \rho_{m, u, \alpha} \binom{m}{j} \epsilon^j (1-\epsilon)^{m-j}\right)\nonumber \\
&\quad \times \left( \sum_{m = z}^{d_c} \rho_{m, x, \alpha} \binom{m}{z} \epsilon^z (1-\epsilon)^{m-z}\right).
\end{align}

\begin{rema}
If $\alpha = 1$, the above initial condition coincides with that for the SC-LDPC ensemble in \cite{olmos}. Therefore, this is a natural generalization of it.
\end{rema}

\section{Prediction of the decoding error probability of the SFC-LDPC ensemble}
In this section, we combine the solution of the EGE for SFC-LDPC ensemble\cite{SFC_EGE} with the solution of the CE for SFC-LDPC ensemble derived in the preceding section in order to predict the finite-length decoding error probability of the SFC-LDPC ensemble.

\begin{figure}[tb]
\centering
\includegraphics[width=3in]{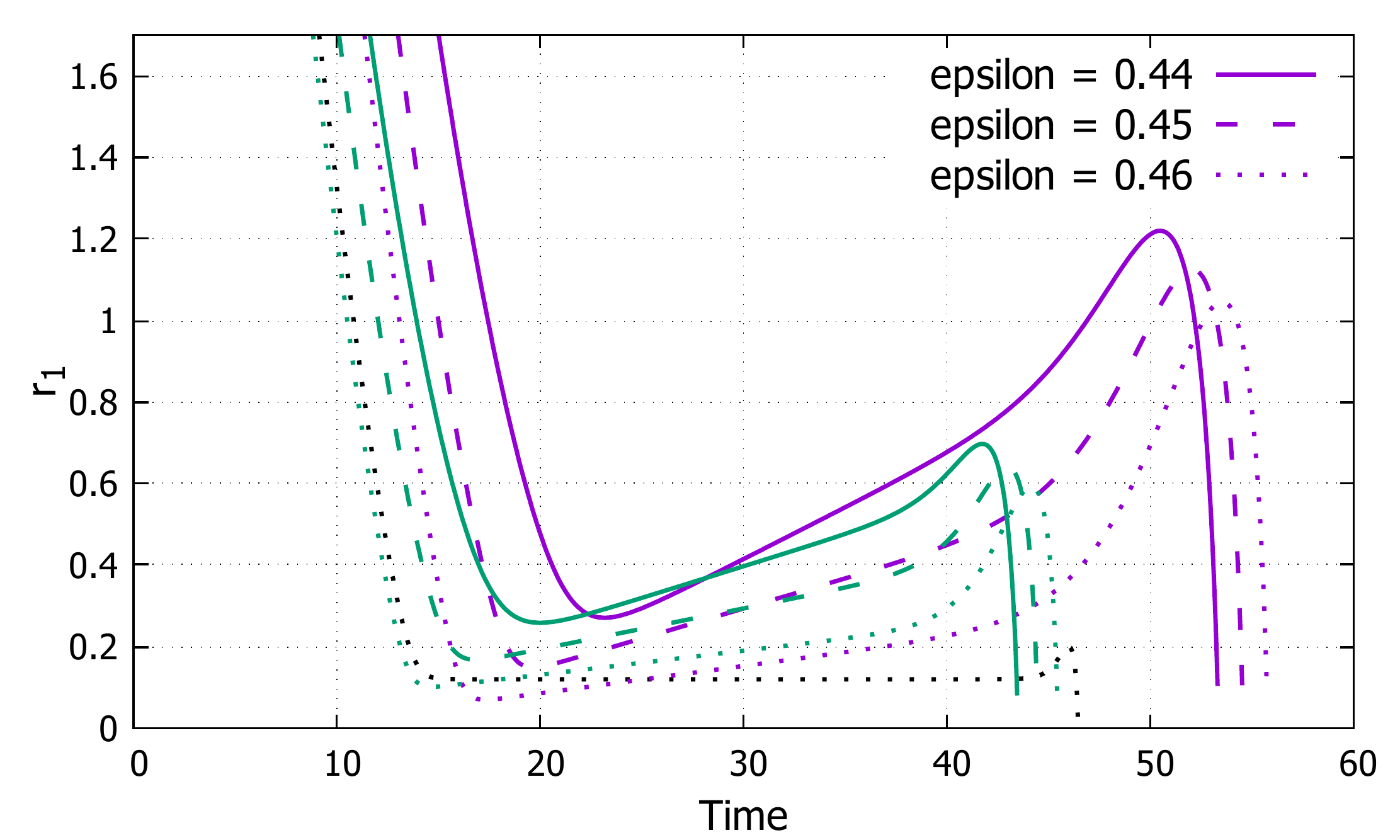}
\caption{$\hat{r}_1(\tau)$ for the (3,6,20,1.1) SFC-LDPC ensemble (the purple lines) and the (3,6,25,1.05) SFC-LDPC ensemble (the green lines) with $\epsilon = 0.44, 0.45, 0.46$ from upper. The dotted black line is for (3,6,50) SC-LDPC ensemble with $\epsilon = 0.46$ for comparison.}
\label{ege}
\end{figure}

\begin{figure}[tb]
\centering
\includegraphics[width=3in]{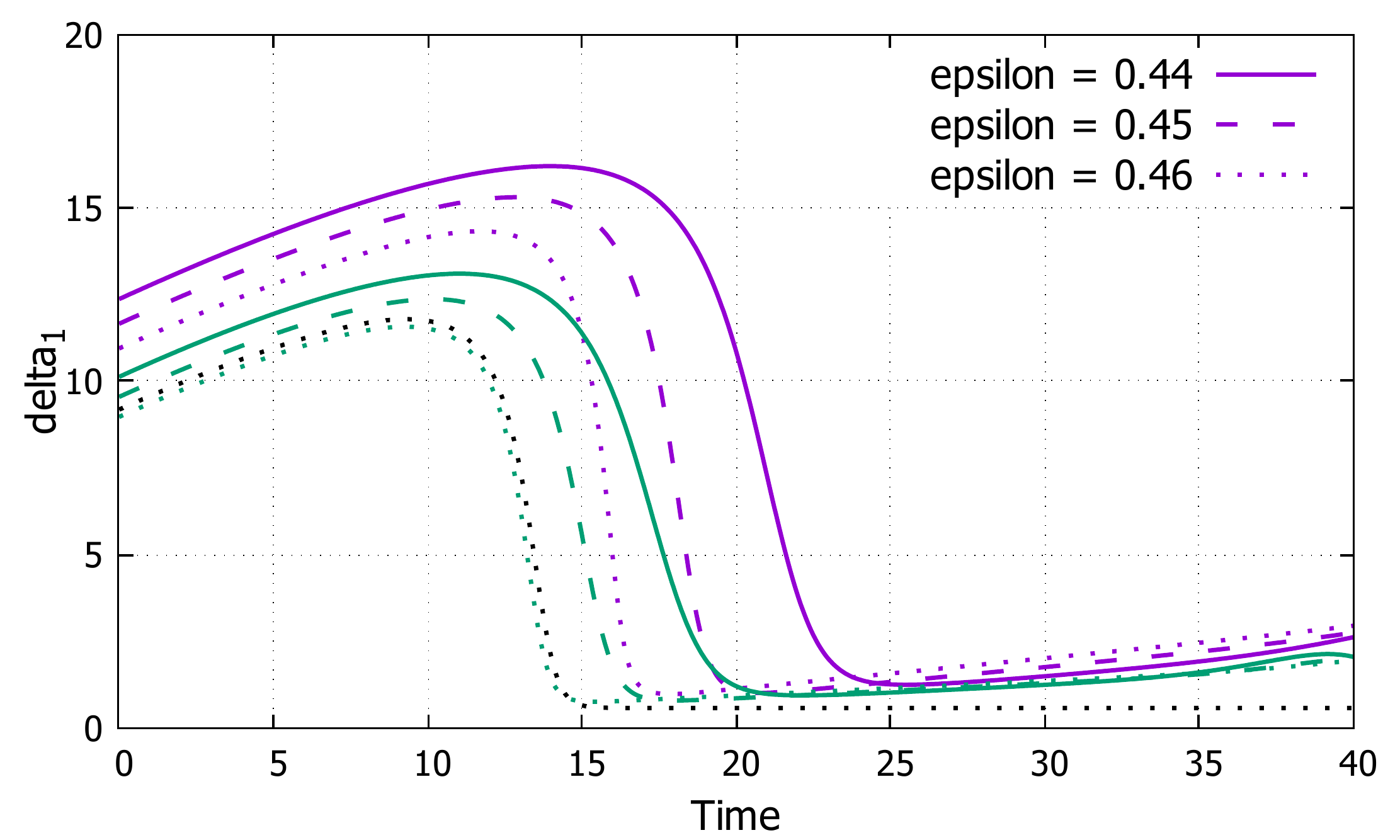}
\caption{$\delta_1(\tau)$ for the (3,6,20,1.1) SFC-LDPC ensemble (the purple lines) and the  (3,6,25,1.05) SFC-LDPC ensemble (the green lines) with $\epsilon = 0.44, 0.45, 0.46$ from upper. The dotted black line is for (3,6,50) SC-LDPC ensemble with $\epsilon = 0.46$ for comparison.}
\label{ce}
\end{figure}

We reproduce the average number of the degree 1 check nodes $\hat{r}_{1}(\tau) = \sum_{u=-L}^L \hat{r}_{1, u}(\tau)$ calculated from the solution of the EGE for the SFC-LDPC ensemble in Fig.\ \ref{ege}, which is derived numerically with the classical Runge-Kutta method in \cite{SFC_EGE}. Figure \ref{ce} shows the variance of the number of degree 1 check nodes $\delta_1 (\tau) =$ $\sum_{u=-L}^{L} \sum_{x=-L}^{L}$ $\delta_{1,x}^{1,u}(\tau)$ calculated from the solution of the CE for the SFC-LDPC ensemble, which is derived numerically with the Euler's method. $\hat{r}_1 (\tau)$ and $\delta_1 (\tau)$ have a local minimum, and the smaller $\epsilon$ is, the ``sharper'' $\hat{r}_1 (\tau)$ is around the local minimum. In addition, the larger $\alpha$ is, the ``sharper'' $\hat{r}_1 (\tau)$ is too.

This is a special feature of the SFC-LDPC ensemble because the previous SC-LDPC ensemble does not have such a local minimum but a flat part called a critical phase.
Therefore, they had to regard $r_1 (\tau)$ of the previous SC-LDPC ensemble as an Ornstein-Uhlenbeck process to approximate the block error probability. However, we are able to approximate the block error probability of the SFC-LDPC ensemble by the probability that the error event occurs on the local minimum like the case of block LDPC codes\cite{amraoui}. Note that the error event occurs when $r_1(\tau) = 0$ before all nodes are removed.

Let $\tau^*$ denote the time when $\hat{r}_1 (\tau)$ is the local minimum point. As shown in Fig.\ \ref{ege}, $\hat{r}_1 (\tau^*)$ looks almost proportional to $\epsilon^{\mathrm{BP}} - \epsilon$, where BP thresholds of the (3,6,20,1.1) and (3,6,25,1.05) SFC-LDPC ensemble are 0.4703 and 0.4785, respectively. On the other hand, $\delta_1 (\tau^*)$ is almost constant for $\epsilon$. As mentioned in Section 3, $r_{j, u} (\tau)$ is Gaussian distributed with mean $\hat{r}_{j,u}(\tau)$ and variance $\delta_{j,u}^{j,u}(\tau)/M$ for a sufficiently large $M$\cite{olmos}. Then, we approximate $\hat{r}_1 (\tau^*)$ by $\gamma (\epsilon^\mathrm{BP} - \epsilon)$, and approximate the ensemble average block error probability by
\begin{align}
Q\left(\frac{\gamma(\epsilon^\mathrm{BP} - \epsilon)}{\sqrt{\delta_1(\tau^*)/M}}\right), \label{Q}
\end{align}
where $Q(\cdot)$ denotes the Q-function, and $\gamma$ is calculated by $\gamma = \frac{r_1(\tau^*)|_{\epsilon = \epsilon^\mathrm{BP}-0.01}}{\epsilon^\mathrm{BP}-(\epsilon^{\mathrm{BP}}-0.01)}$ for each $L$ and $\alpha$. Therefore, we consider that the BP threshold $\epsilon^{\mathrm{BP}}$ affects the ``position'' of the waterfall and the coefficient $\gamma / \sqrt{\delta_1 (\tau^*)}$ affects the ``steepness'' of the waterfall. Table \ref{gammadelta1} shows $\epsilon^{\mathrm{BP}}$, $\gamma$, $\delta_1 (\tau^*)$, and $\gamma / \sqrt{\delta_1 (\tau^*)}$ of $(3,6,20,\alpha)$ SFC-LDPC ensemble for several $\alpha$. Although these parameters are depend on $L$ strictly speaking, it is observed that these parameters are almost independent of $L$. The actual parameters for the ensembles used in our experiments are shown in the next section.

Then, we expect that the larger $\alpha$ is, the steeper the waterfall is and the more left-shifted. Note that $M$ also increases as $\alpha$ increases under the fixed code length and the design rate, as described in Remark \ref{alphaml}.

\begin{table}[tb]
\caption{$\epsilon^{\mathrm{BP}}$, $\gamma$, $\delta_1(\tau^*)$ and $\gamma / \sqrt{\delta_1 (\tau^*)}$ for $(3, 6, 20, \alpha)$ SFC-LDPC ensemble}
\label{gammadelta1}
\centering
\begin{tabular}{c | c c c c}
\hline
$\alpha$ &$\epsilon^\mathrm{BP}$&$\gamma$ & $\delta_1(\tau^*)$ & $\gamma / \sqrt{\delta_1(\tau^*)}$\\
\hline
1.05 & 0.4785 & 5.39 & 0.806 & 6.00\\
1.10 & 0.4703 & 6.77 & 1.03 & 6.67\\
1.15 & 0.4631 & 8.68 & 1.39 & 7.36\\
1.20 & 0.4571 & 11.6 & 2.12 & 7.95\\
\hline
\end{tabular}
\end{table}

\section{Experiments}

\subsection{Experiment conditions}
The parameters of the ensembles used in the experiments are shown in Table \ref{ensembles} and \ref{gammadelta}. We generate 1100 codes and 1000 codewords from each code for A1--A4 with $\epsilon \leq 0.440$, B1 with $\epsilon \leq 0.450$, and C1 with $\epsilon \leq 0.455$, and we generate 100 codes and 1000 codewords from each code for A1--A4 with $\epsilon \geq 0.445$, B1 with $\epsilon \geq 0.455$, and C1 with $\epsilon \geq 0.460$. The rates shown in Table \ref{ensembles} are the average rate of those generated codes. The decoder is the BP decoder with no limitation of the number of iterations, which has the same decoding error probability as that of the peeling decoder. Note that we remove the small cycles of Tanner graphs, whose lengths are lower than or equal to 6, in order to observe the block error probability in the waterfall region more precisely. 

\begin{table}[tb]
\caption{The parameters used in the experiments}
\label{ensembles}
\centering
\begin{tabular}{c | c c c c c c}
\hline
 & $(d_v, d_c)$ & $\alpha$ & $L$ & $M$& Length & Rate\\
\hline
A1 &(3, 6)&1.1&7&500&10,469&0.460\\
A2 &(3, 6)&1.1&10&500&17,243&0.476\\
A3 &(3, 6)&1.1&15&500&33,875&0.487\\
A4 &(3, 6)&1.1&20&500&60,656&0.493\\
\hline
B1 &(3, 6)&1.1&10&1000&34,478&0.476\\
\hline
C1 &(3, 6)&1.05&10&1000&26,795&0.467\\
\hline
\end{tabular}
\end{table}


\begin{table}[tb]
\caption{Parameters in (\ref{Q}) for the ensembles used in the experiments}
\label{gammadelta}
\centering
\begin{tabular}{c | c c c c}
\hline
 &$\epsilon^\mathrm{BP}$&$\gamma$ & $\delta_1(\tau^*)$ & $\gamma / \sqrt{\delta_1(\tau^*)}$\\
\hline
A1 &0.4710&6.70&1.09&6.41\\
A2 &0.4703&6.77&1.03&6.68\\
A3 &0.4703&6.76&1.03&6.67\\
A4 &0.4703&6.77&1.03&6.67\\
\hline
B1 &0.4703&6.77&1.03&6.67\\
\hline
C1 &0.4785&5.39&0.807&6.00\\
\hline
\end{tabular}
\end{table}

\subsection{Difference in $M$}
\begin{figure}[tb]
\centering
\includegraphics[width=3in]{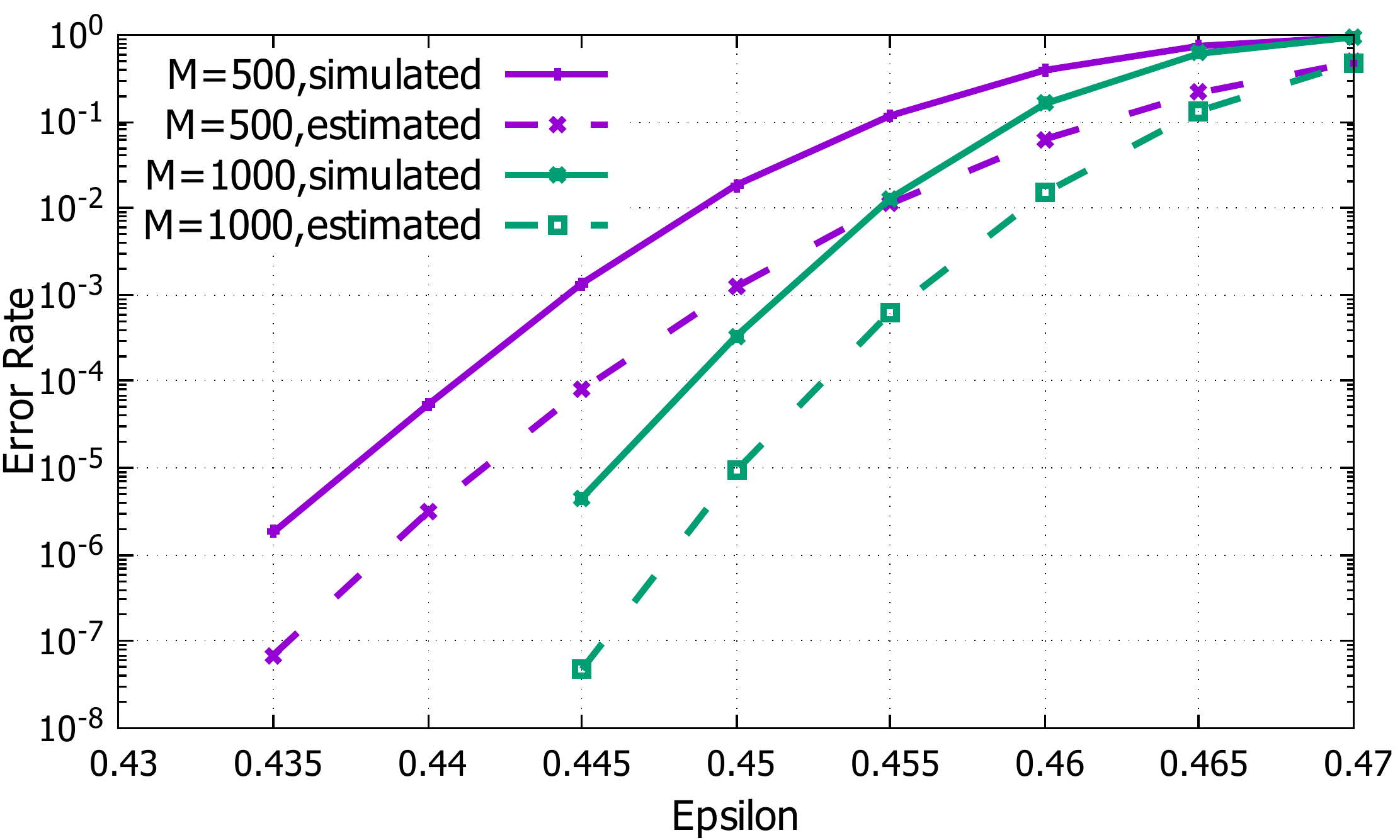}
\caption{Simulated block error probability curves and the estimated curves for ensembles A2 and B1 in Table \ref{ensembles}}
\label{M}
\end{figure}
Figure \ref{M} shows that the simulated block error probability curves and the estimated curves for several $M$. The estimated curves approximate the simulated curves except for the difference in some shift on the semilog graph. The larger $M$ become, the smaller the horizontal sifted width between the simulated curve and the estimated curve becomes.

\subsection{Difference in $L$}
\begin{figure}[tb]
\centering
\includegraphics[width=3in]{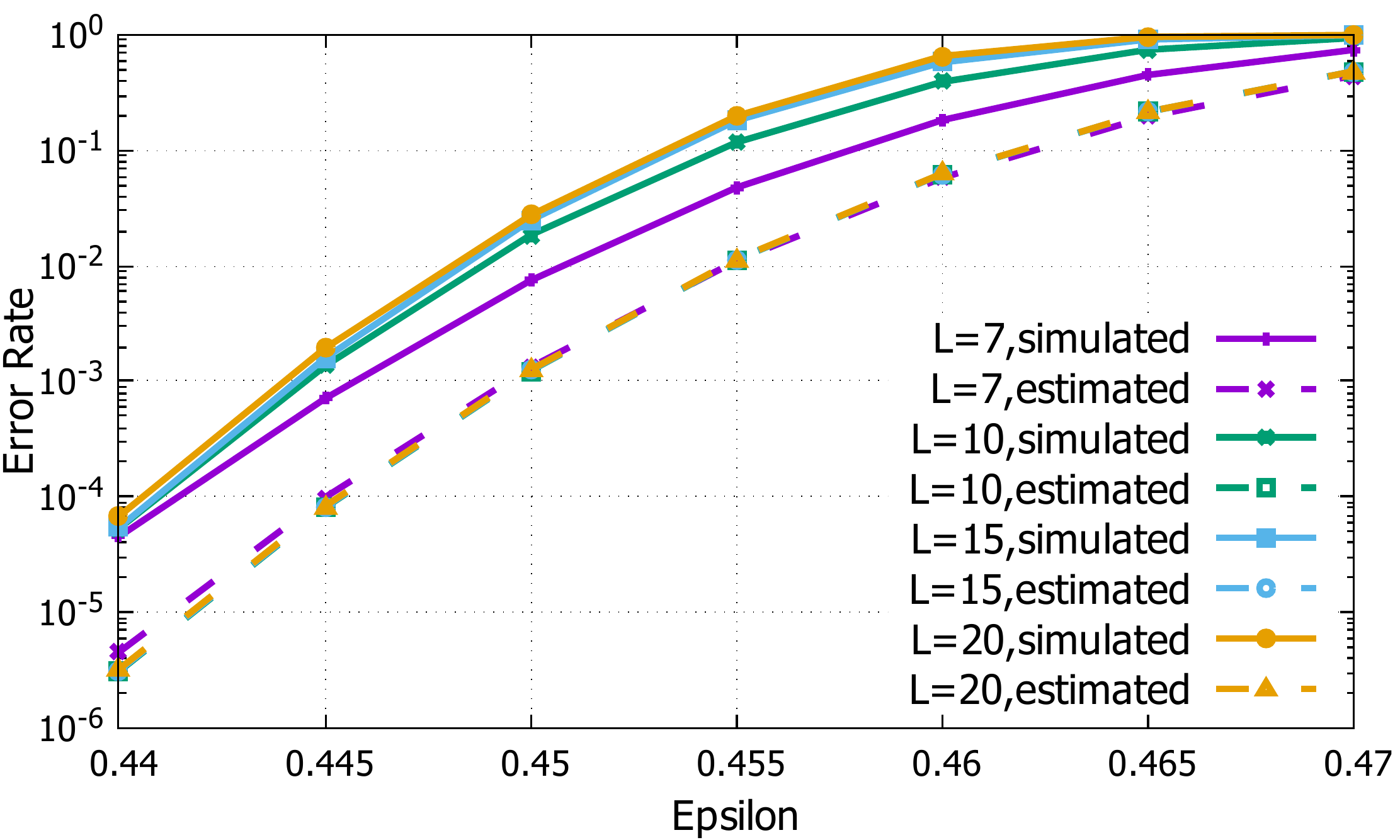}
\caption{Simulated block error probability curves and the estimated curves for ensembles A1, A2, A3, and A4 in Table \ref{ensembles}}
\label{L}
\end{figure}
Figure \ref{L} shows that the simulated block error probability curves and the estimated curves for several $L$. Simulated block error probability is hoped to be independent of $L$, because the approximated block error probability (\ref{Q}) does not depend on $L$. Unfortunately, Fig.\ \ref{L} shows that the larger $L$ is, the larger block error probability is. However, Fig.\ \ref{L} also shows that the amount of increase of the block error probability decreases as $L$ increases. In particular, the amount of increase of block error probability is small when $\epsilon$ is small. This should be because the smaller $\epsilon$ is, the sharper the graph around the local minimum is, as shown in Fig.\ \ref{ege}. This phenomenon is not observed for the SC-LDPC codes whose block error probability increases proportionally to $L$ in a wide range of $\epsilon$.

\subsection{Difference in $\alpha$}
\begin{figure}[tb]
\centering
\includegraphics[width=3in]{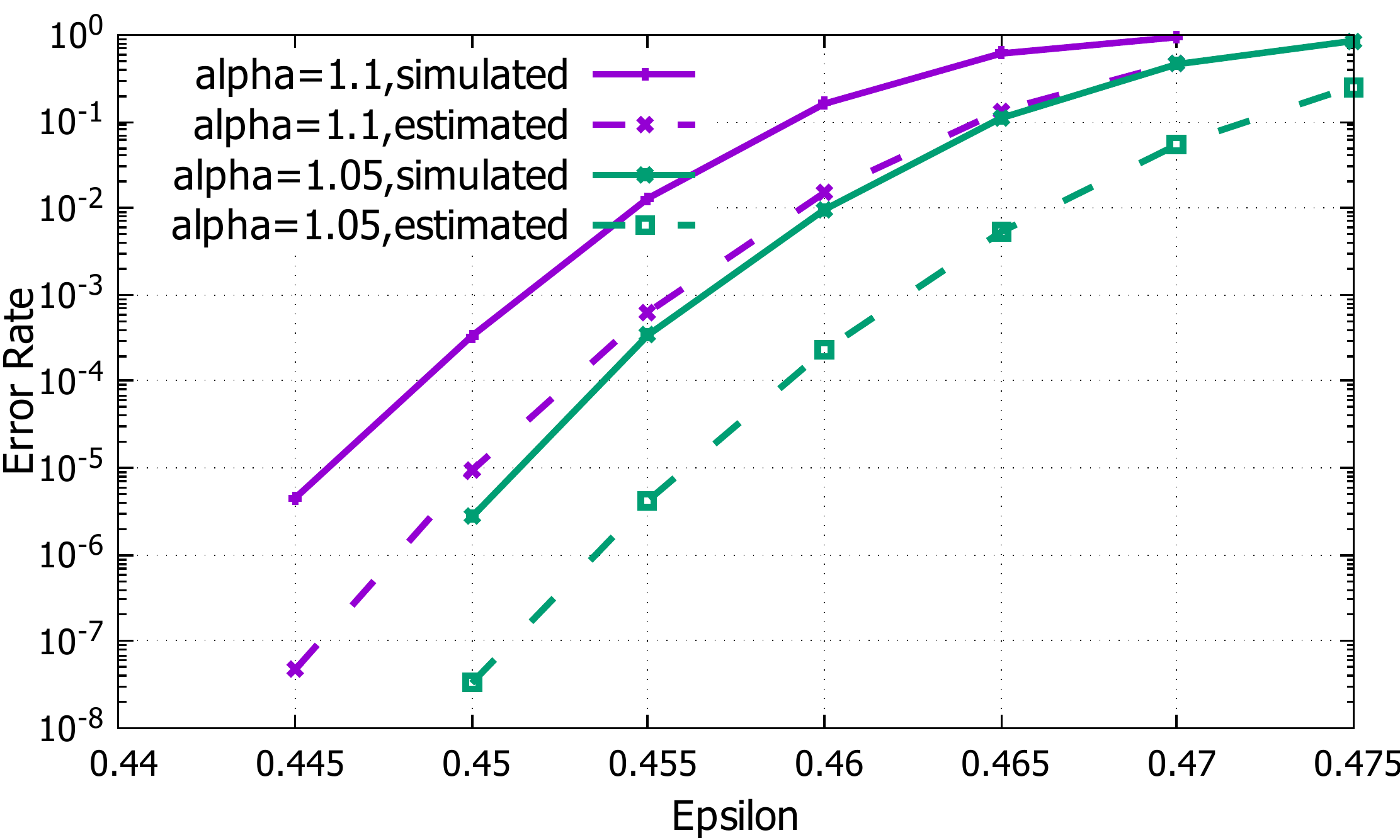}
\caption{Simulated block error probability curves and the estimated curves for ensembles B1 and C1 in Table \ref{ensembles}}
\label{alpha}
\end{figure}
Figure \ref{alpha} shows that block error probability curves and estimated curves for several $\alpha$. Figure \ref{alpha} shows that the larger $\alpha$ is, the better the approximation is. One of the  reason should be because the larger $\alpha$ is, the sharper the graph of $\hat{r}_1 (\tau)$ around the local minimum is, as shown in Fig.\ \ref{ege}.

\subsection{Code construction}
From the preceding analysis, we expect that the decoding error probability in the waterfall region over the BEC is left-shifted and becomes steeper as $\alpha$ increases. In addition, it has already been observed in \cite{SFC} that the average number of iterations of the SFC-LDPC ensemble in the waterfall region is less than that of the SC-LDPC ensemble. Moreover, it has been analyzed in \cite{SFC_weight_distribution} that the decoding error probability in the error floor region over the BEC is lower than that of the SC-LDPC ensemble.

Then, tuning $\alpha$ appropriately, we can construct an SFC-LDPC ensemble with the following 3 properties.
1. It has the same rate and code length as those of the target SC-LDPC ensemble. 2. It has a lower decoding error probability than the target ensemble under the condition $\epsilon < \epsilon^*$, where $\epsilon^*$ is a target channel erasure probability. 3. It has a lower average number of iterations than that of the target ensemble.

\begin{figure}[tb]
\centering
\includegraphics[width=3in]{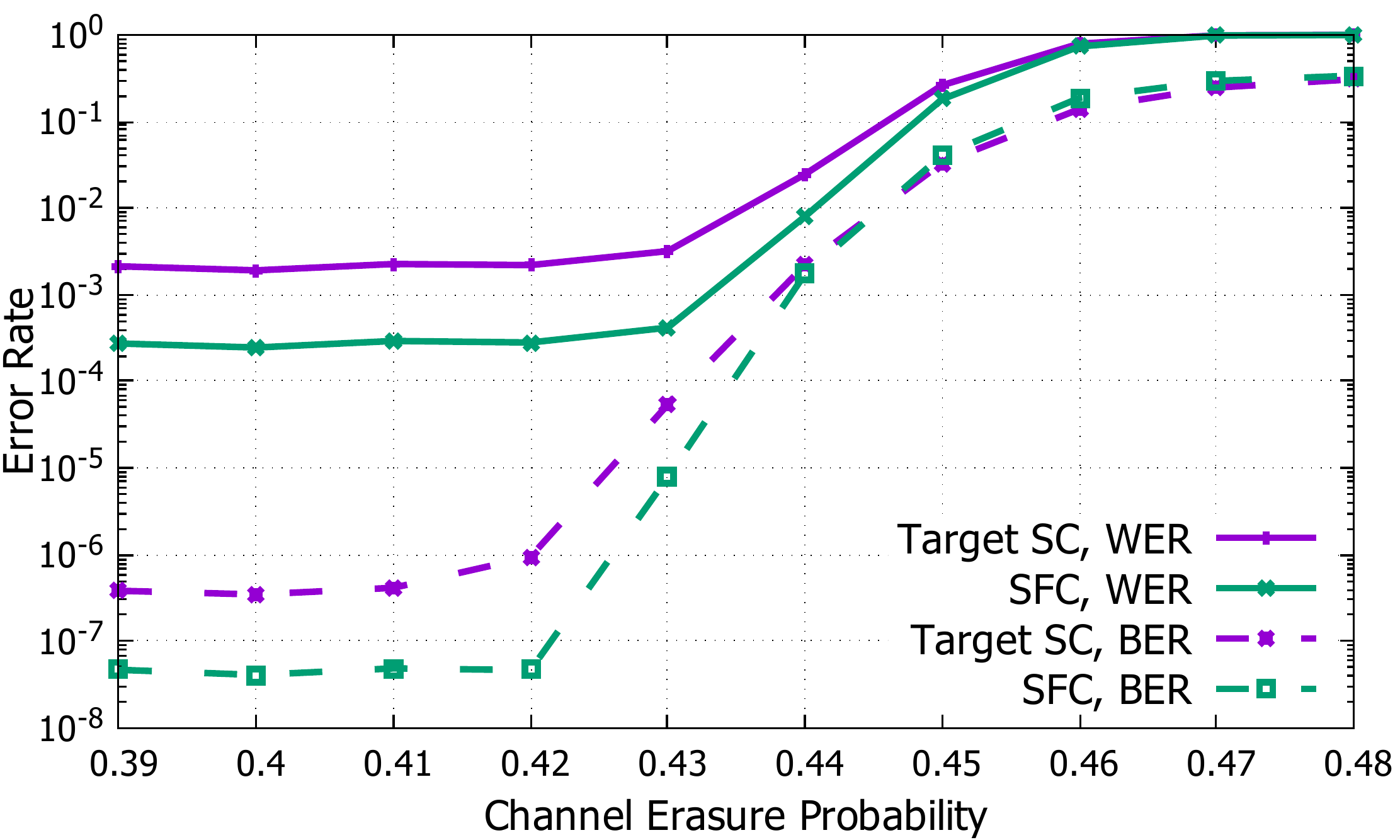}
\caption{Word error rate and bit error rate of the target SC-LDPC ensemble and the constructed SFC-LDPC ensemble on the BEC.}
\label{SCSFC}
\end{figure}

\begin{figure}[tb]
\centering
\includegraphics[width=3in]{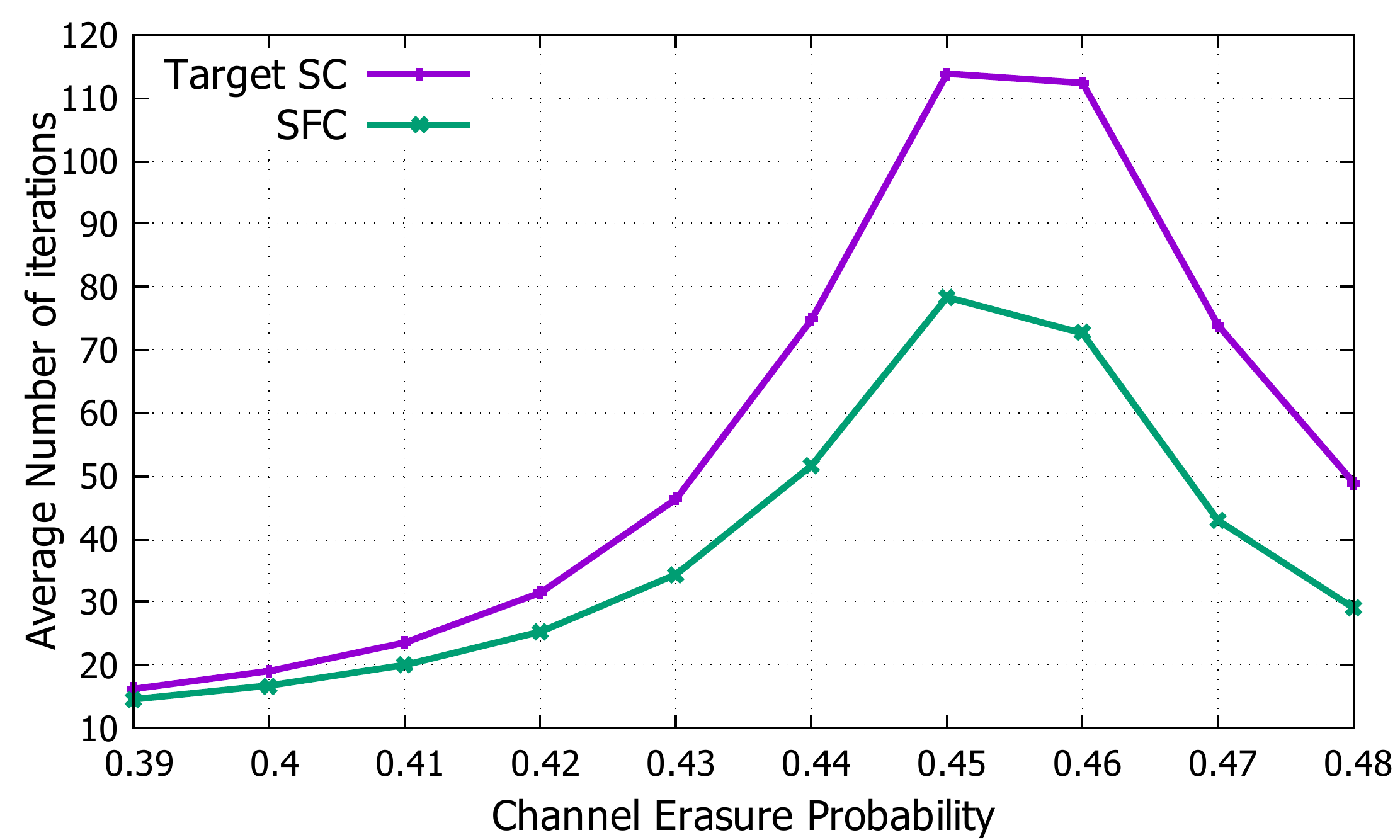}
\caption{Average number of iterations of the target SC-LDPC ensemble and the constructed SFC-LDPC ensemble on the BEC.}
\label{SCSFCit}
\end{figure}

Actually, we construct an SFC-LDPC ensemble with the above properties. The parameters of the target SC-LDPC ensemble and the constructed SFC-LDPC ensemble are in Table \ref{ensemble_parameters}. The small cycles of Tanner graphs, whose lengths are lower than or equal to 6 are removed. Target channel erasure probability is $\epsilon^* = 0.44$. Figure \ref{SCSFC} shows the decoding error probability of those ensembles derived by Monte Carlo simulation. When $\epsilon \leq 0.42$, 11000 codes and 100 codewords from each code are generated. When $\epsilon > 0.42$,  1000 codes and 100 codewords from each code are generated. Figure \ref{SCSFCit} shows the average number of iterations of them. The constructed ensemble has the desired properties.

\begin{figure}[tb]
\centering
\includegraphics[width=3in]{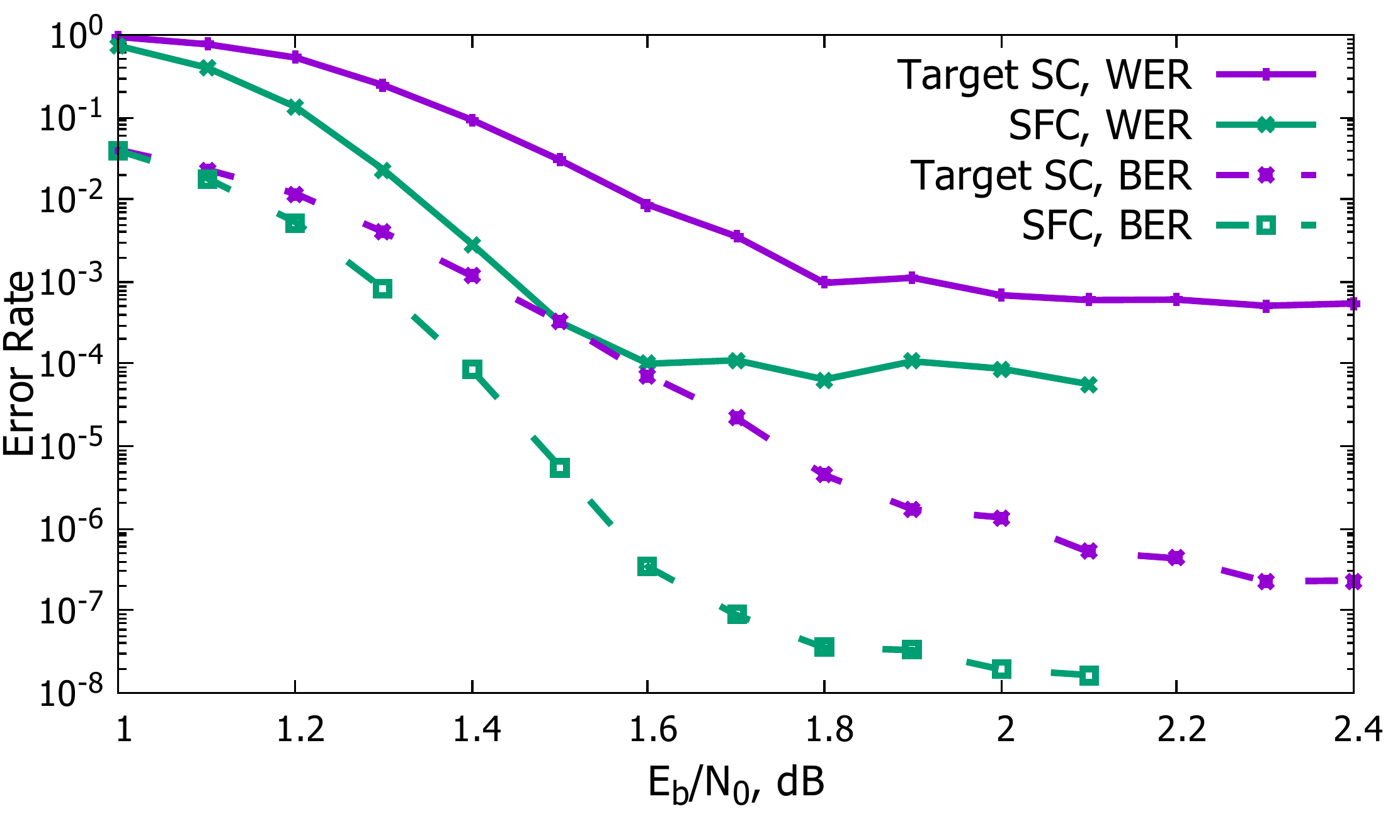}
\caption{Word error rate and bit error rate of the target SC-LDPC ensemble and the constructed SFC-LDPC ensemble on the AWGN channel.}
\label{SCSFCawgn}
\end{figure}

\begin{figure}[tb]
\centering
\includegraphics[width=3in]{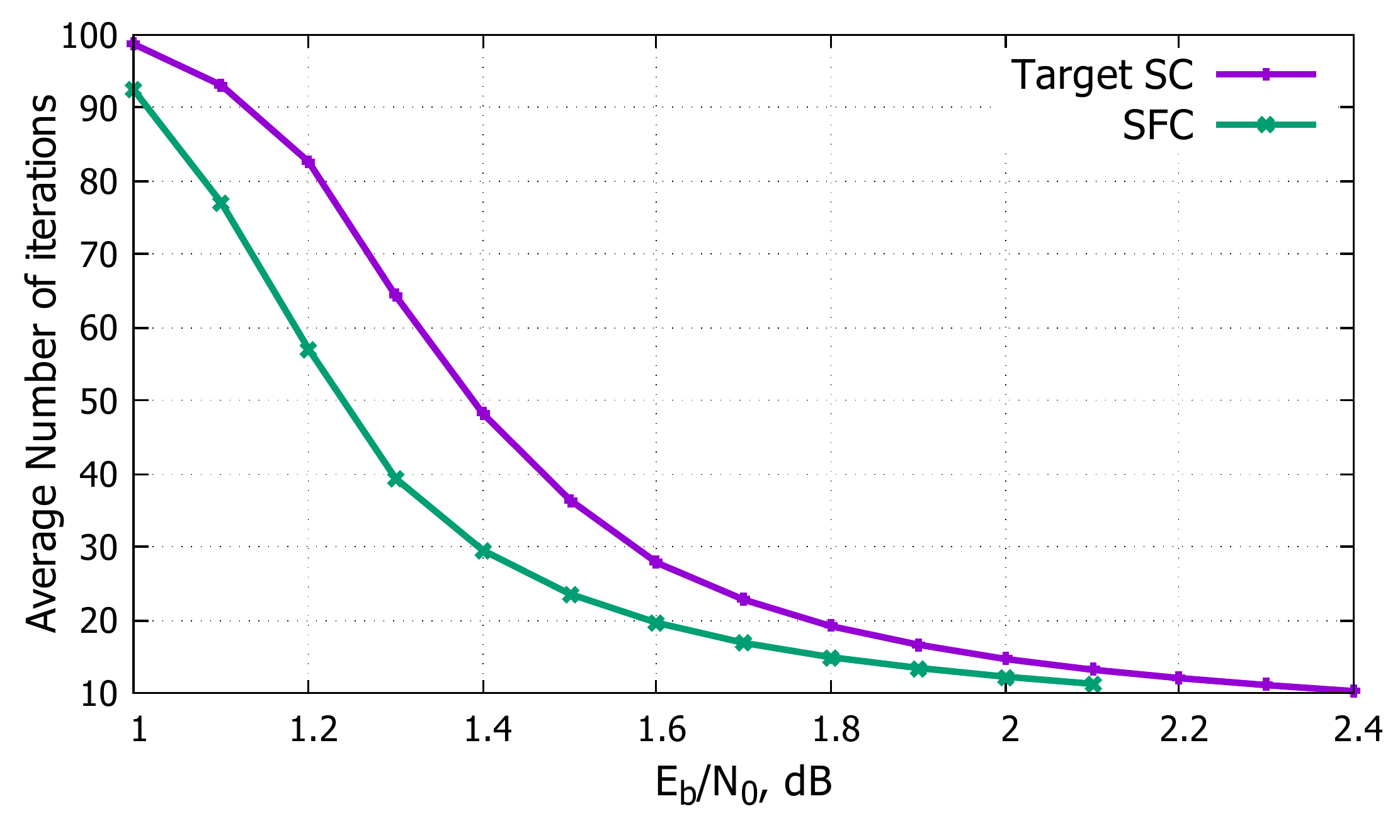}
\caption{Average number of iterations of the target SC-LDPC ensemble and the constructed SFC-LDPC ensemble on the AWGN channel.}
\label{SCSFCit_awgn}
\end{figure}

As a supplement, simulation results over the AWGN channels are shown in Fig.\ \ref{SCSFCawgn} and \ref{SCSFCit_awgn} for practical interest. From target SC-LDPC ensemble, we generate 1000 codes and 100 codewords when $E_b/N_0 < 2.0$, and 10000 codes and 100 codewords from each code when $E_b/N_0 \geq 2.0$. From constructed SFC-LDPC ensemble, we generate 1000 codes and 100 codewords when $E_b/N_0 < 1.5$, and 10000 codes and 100 codewords from each code when $E_b/N_0 \geq 1.5$. Maximum number of iterations is set at 100. It shows that the constructed ensemble has similar properties on the BEC.

\begin{table}[!t]
\caption{The parameters of the target SC-LDPC ensemble and the constructed SFC-LDPC ensemble}
\label{ensemble_parameters}
\centering
\begin{tabular}{c | c c c c c c c c}
\hline
& $d_v$ & $d_c$ & $L$ & $\alpha$ & $M$ & Code length & Rate\\
\hline
Target & 3 & 6 & 25 & 1.00 & 250 & 12,750 & 0.482\\
SFC & 3 & 6 & 12 & 1.11 & 260 & 12,734 & 0.483\\
\hline
\end{tabular}
\end{table}

\section{Conclusion}
In this paper, we derived the CE for the SFC-LDPC ensemble as the last piece of the theoretical analysis over the BEC. We combined its solution with the solution of the EGE for the SFC-LDPC ensemble, which had been derived in \cite{SFC_EGE}. Then, we analyzed the decoding error probability of the SFC-LDPC ensemble in the waterfall region. The waterfall became steeper as $\alpha$ increased. As a result, it mitigated the decrease of the BP threshold.

\appendices
\section{}\label{A}
For $u \in [-L, L+d_v-1], j \in [1, d_c]$, the initial conditions of the EGE for SFC-LDPC ensemble are as follows.
\begin{align}
\hat{r}_{j,u} (0) \!=\! j \frac{1}{d_c} \! \left( \sum_{k=0}^{d_v\!-\!1} \alpha^{L\!-\!|u\!-\!k|} \right) \! \sum_{m\geq j}^{d_c} \rho_{m, u, \alpha} \binom{m}{j} \epsilon^j (1\!-\!\epsilon)^{m\!-\!j},
\end{align}
where $\rho_{m,u, \alpha}$ is defined in (\ref{lem1-5}).
\begin{align}
\hat{v}_{u}(0) = 
\begin{cases}
\epsilon \alpha^{L-|u|}, & u \in [-L,L],\\
0, & \mathrm{otherwise}.
\end{cases}
\end{align}

\section{}\label{B}
Our proof mainly follows the Appendix C of \cite{olmos} but more detailed and generalized. In addition to the notation in Section \ref{sec3}, let $\delta_{z, x}^{j, u} (\tau)$ also denote $\mathrm{CoVar}[V_u (t M), R_{z, x} (t M)] / M = \mathrm{CoVar}[v_u (\tau), r_{z, x} (\tau)] M$ for $j = d_c+1, z \leq d_c$ and $\mathrm{CoVar}[V_u (t M), V_x (t M)] / M = \mathrm{CoVar}[v_u (\tau), v_x (\tau)] M$ for $j = z = d_c +1$.
Initial conditions of the covariance evolution are as follows. Let $p_{j,u,\alpha}$ denote the probability that a randomly chosen check node at position $u$ has the degree $j$ after the peeling decoder initialization. It is given by
\begin{align}
p_{j,u, \alpha} = \sum_{m=j}^{d_c} \rho_{m,u,\alpha} \binom{m}{j} \epsilon^j (1-\epsilon)^{m-j}.
\end{align}
Initial conditions of the covariance evolution for the SFC-LDPC ensemble are divided into the following three cases in the same manner as those for the SC-LDPC ensemble.
\begin{align}
&\delta^{j,u}_{z,x}(0) \nonumber \\
&=
\begin{cases}
\mathrm{CoVar} [R_{j,u}(0), R_{z,x}(0)] / M, & j, z \leq d_c,\\
\mathrm{CoVar} [V_u(0), R_{z,x}(0)] / M, & j = d_c + 1,\, z \leq d_c,\\
\mathrm{CoVar} [V_u(0), V_x(0)] / M, & j = z = d_c + 1
\end{cases}
\end{align}

When $j = z = d_c + 1$, $V_u(0)$ follows a binomial distribution with $\alpha^{L-|u|} M$ trials and probability $\epsilon$ independently from each other position. Therefore, 
\begin{align}
\delta^{d_c+1,u}_{d_c+1, x}(0) &= \mathrm{CoVar} [V_u (0), V_x(0)] / M \nonumber \\
&= 
\begin{cases}
\mathrm{Var}[V_u(0)] / M = \alpha^{L-|u|} \epsilon (1-\epsilon), & u = x\\
0, & u \neq x
\end{cases}\label{ce_gsc_begin1}
\end{align}

When $j, z \leq d_c$, we divide the cases as follows.
\begin{align}
\begin{cases}
u = x
\begin{cases}
j = z\\
j \neq z
\end{cases}\\
u \neq x
\begin{cases}
|u-x| \geq d_v\\
|u-x| < d_v
\end{cases}
\end{cases}\nonumber
\end{align}

For $u = x$, the number of degree $j$ check nodes at position $u$ follows a multinomial distribution with $\frac{1}{d_c} M \sum_{k=0}^{d_v-1} \alpha^{L-|u-k|}$ trials and probability $p_{j,u,\alpha}$. Therefore, for $u = x$, $j = z$,
\begin{align}
\delta^{j,u}_{j,u}(0) &= \mathrm{Var}[R_{j,u}(0)] / M \\
&= j^2 \frac{1}{d_c} \left( \sum_{k=0}^{d_v-1}\alpha^{L-|u-k|} \right) p_{j, u, \alpha} (1-p_{j,u,\alpha}),
\end{align}
and for $u=x$ and $j \neq z$,
\begin{align}
\delta^{j,u}_{z,u}(0) &= \mathrm{CoVar}[R_{j,u}(0), R_{z,u}(0)] /M \\
&= -jz \frac{1}{d_c} \left( \sum_{k=0}^{d_v-1}\alpha^{L-|u-k|} \right) p_{j,u,\alpha} p_{z,u,\alpha}.
\end{align}

For $u \neq x$ and  $|u-x| \geq d_v$, any check node at position $u$ and any check node at position $x$ cannot be connected to each other by one variable node, and they are independent trough the peeling decoder initialization. Therefore,
\begin{align}
\delta^{j,u}_{z,x}(0) = \mathrm{CoVar}[R_{j,u}(0), R_{z,x}(0)] / M = 0. \label{ce_gsc_end1}
\end{align}

For $u \neq x$ and $|u-x| < d_v$, we assume $x > u$ without loss of generality. In this case, we have to consider the effect from a check node at position $u$ and a check node at position $x$ which share at least one variable node before the dummy node shortening and the initialization of the peeling decoder. Let $\mathsf{check}_u$ and $\mathsf{check}_x$ denote a pair of check nodes selected at random from positions $u$ and $x$, respectively. There are $d_v - |u-x|$ positions, from $x-d_v+1$ to $u$, in which any variable node is connected with one edge to a check node at position $u$ and with one edge to a check node at position $x$. $\mathsf{check}_u$ has $a$ edges connected to variable nodes at positions $[x-d_v+1, u]$, and the number $a$ is according to a binomial distribution with $d_c$ trials with probability
\begin{align}
\frac{\sum_{k=x-d_v+1}^u \alpha^{L-|k|}}{\sum_{j=0}^{d_v-1}\alpha^{L-|u-j|}}.
\end{align}
$\mathsf{check}_x$ has $b$ edges in the same manner as $\mathsf{check}_u$. Note that $a$ and $b$ are independent random variables.

For a given pair $(a, b)$, the probability that $\mathsf{check}_u$ and $\mathsf{check}_x$ share at least one variable node at positions $[x-d_v+1, u]$ is approximated as follows, where the sampling without replacement is approximated by the sampling with replacement.
\begin{align}
1 - \left( \frac{\sum_{k=x\!-\!d_v\!+\!1}^u \alpha^{L\!-\!|k|} M - a}{\sum_{k=x\!-\!d_v\!+\!1}^u \alpha^{L\!-\!|k|} M} \right)^b
\sim \frac{ab}{\sum_{k=x\!-\!d_v\!+\!1}^u \alpha^{L\!-\!|k|} M},\label{ab/X}
\end{align}
for sufficiently large $M$ by ignoring the terms $O(M^{-2})$. We ignore the case that $\mathsf{check}_u$ and $\mathsf{check}_x$ share two or more variable nodes since the probability of such a case decays by $O(M^{-2})$. Then, averaging (\ref{ab/X}) over all possible pairs $(a,b)$, we can evaluate the probability $P_S$ that $\mathsf{check}_u$ and $\mathsf{check}_x$ share at least one variable node before dummy node shortening by
\begin{align}
P_S &= \frac{1}{\sum_{k=x-d_v+1}^u \alpha^{L-|k|} M} \cdot d_c \left( \frac{\sum_{k=x-d_v+1}^u \alpha^{L-|k|}}{\sum_{j=0}^{d_v-1}\alpha^{L-|u-j|}} \right) \nonumber \\
& \quad \cdot d_c \left( \frac{\sum_{k=x-d_v+1}^u \alpha^{L-|k|}}{\sum_{j=0}^{d_v-1}\alpha^{L-|x-j|}} \right) \\
&= \frac{d_c^2 \sum_{k=x-d_v+1}^u \alpha^{L-|k|}}{M \left( \sum_{j=0}^{d_v-1}\alpha^{L-|u-j|} \right) \left( \sum_{j=0}^{d_v-1}\alpha^{L-|x-j|} \right)}.
\end{align}

Because the probability that the shared variable node is not a dummy node is
\begin{align}
\frac{\sum_{l = \max \{ -L, x-d_v-1\}}^{\min \{ L, u\}} \alpha^{L-|l|}}{\sum_{k=x-d_c+1}^{u} \alpha^{L-|k|}},
\end{align}
the probability $P'_S$ that $\mathsf{check}_u$ and $\mathsf{check}_x$ share at least one variable node after dummy node shortening before the peeling decoder initialization is
\begin{align}
P'_S &=  \frac{\sum_{l = \max \{ -L, x-d_v-1\}}^{\min \{ L, u\}} \alpha^{L-|l|}}{\sum_{k=x-d_c+1}^{u} \alpha^{L-|k|}} \nonumber \\
&\quad \cdot \frac{d_c^2 \sum_{k=x-d_v+1}^u \alpha^{L-|k|}}{M \left( \sum_{j=0}^{d_v-1}\alpha^{L-|u-j|} \right) \left( \sum_{j=0}^{d_v-1}\alpha^{L-|x-j|} \right)}\\
&= \frac{d_c^2 \sum_{l = \max \{ -L, x-d_v-1\}}^{\min \{ L, u\}} \alpha^{L-|l|}}{M \left( \sum_{j=0}^{d_v-1}\alpha^{L-|u-j|} \right) \left( \sum_{j=0}^{d_v-1}\alpha^{L-|x-j|} \right)}
\end{align}

Let $d_u$ and $d_x$ denote the degree of $\mathsf{check}_u$ and $\mathsf{check}_x$ after the peeling decoder initialization, respectively. Then, the probability that $P(d_u = j, d_x = z)$ can be expressed as follows.
\begin{align}
&P(d_u=j, d_x=z) \nonumber \\
&= P(d_u=j,d_x=z | \mathrm{share}) P'_S \nonumber \\
&\quad + P(d_u=j,d_x=z | \mathrm{no \ share}) (1-P'_S),
\end{align}
where $P(d_u=j,d_x=z | \mathrm{share})$ denotes the conditional probability that $\mathsf{check}_u$ and $\mathsf{check}_u$ share one variable (not dummy) node. It is obtained by
\begin{align}
P(&d_u = j, d_x = z | \mathrm{share}) =\nonumber \\
& \epsilon \left[ \left( \sum_{m=j}^{d_c} \rho'_{m,u,\alpha} \binom{m-1}{j-1} \epsilon^{j-1} (1-\epsilon)^{m-j} \right) \right. \nonumber \\
& \qquad \left. \times \left( \sum_{m = z}^{d_c} \rho'_{m, x,\alpha} \binom{m-1}{z-1} \epsilon^{z-1} (1-\epsilon)^{m-z} \right) \right] \nonumber \\
& \quad + (1- \epsilon) \left[ \left( \sum_{m = j+1}^{d_c} \rho'_{m,u,\alpha} \binom{m-1}{j} \epsilon^j (1- \epsilon)^{m-j-1} \right) \right. \nonumber \\
& \qquad \left. \times \left( \sum_{m = z+1}^{d_c} \rho'_{m,x,\alpha} \binom{m-1}{z} \epsilon^z (1-\epsilon)^{m-z-1} \right) \right],
\end{align}
where
\begin{align}
&\rho'_{m,u, \alpha} = \nonumber \\
&\begin{cases}
1, \quad u \in [-L+d_v-1, L], m=d_c, \\
0, \quad u \in [-L+d_v-1, L], m < d_c, \\
\binom{d_c-1}{m-1} \left( s_{u,\alpha} \right)^{m-1} \left( 1 - s_{u, \alpha} \right)^{d_c-m},\\
\quad u \in [-L, -L+d_v-2] \cup [L+1, L+d_v-1].
\end{cases}
\end{align}
The first term of $P(d_u=j,d_x=z | \mathrm{share})$ represents the probability that the shared variable node is erased, and the second term represents the probability that the shared variable node is not erased. In addition, the following holds directly.
\begin{align}
P(d_u = j, d_x = z | \mathrm{no \ share}) = p_{j,u,\alpha}p_{z,x,\alpha}.
\end{align}

Then, $\mathrm{CoVar}[R_{j,u}(0), R_{z,x}(0)]$ is obtained by the following calculation.
\begin{align}
&\mathrm{CoVar}[R_{j,u}(0), R_{z,x}(0)] \nonumber \\
&\quad = \mathbb{E}[R_{j,u}(0) R_{z,x}(0)] - \mathbb{E}[R_{j,u}(0)] \mathbb{E}[R_{z,x}(0)]\\
&\quad = j \left( \frac{M}{d_c} \sum_{k=0}^{d_v-1} \alpha^{L-|u-k|} \right) \cdot z \left( \frac{M}{d_c} \sum_{k=0}^{d_v-1} \alpha^{L-|x-k|} \right) \nonumber \\
&\qquad \cdot P(d_u = j, d_x = z) \nonumber \\
&\qquad - j \left( \frac{M}{d_c} \sum_{k=0}^{d_v-1} \alpha^{L-|u-k|} \right) p_{j,u,\alpha} \nonumber \\
&\qquad \cdot z \left( \frac{M}{d_c} \sum_{k=0}^{d_v-1} \alpha^{L-|x-k|} \right) p_{z,x,\alpha}
\end{align}
\begin{align}
&\quad = j z \left( \frac{M}{d_c} \sum_{k=0}^{d_v-1} \alpha^{L-|u-k|} \right) \left( \frac{M}{d_c} \sum_{k=0}^{d_v-1} \alpha^{L-|x-k|} \right) \nonumber \\
&\qquad \cdot \Biggl\{ P'_S \Bigl( P(d_u = j, d_x = z | \mathrm{share}) \nonumber \\
&\qquad \qquad - P(d_u = j, d_x = z | \mathrm{no \ share}) \Bigr) \nonumber \\
&\qquad \qquad \qquad + p_{j,u,\alpha} p_{z,x,\alpha} \Biggr\} \nonumber \\
&\qquad - j z \left( \frac{M}{d_c} \sum_{k=0}^{d_v-1} \alpha^{L-|u-k|} \right) \left( \frac{M}{d_c} \sum_{k=0}^{d_v-1} \alpha^{L-|x-k|} \right) \nonumber \\
& \qquad \qquad \cdot p_{j,u,\alpha} p_{z,x,\alpha}\\
&\quad = jz M \left( \sum_{k= \max \{-L, x-d_v+1\}}^{\min \{ L, u\}} \alpha^{L-|k|} \right) \nonumber \\
&\qquad \times \Bigl( P(d_u = j, d_x = z | \mathrm{share}) \nonumber \\
&\qquad \qquad - P(d_u = j, d_x = z | \mathrm{no \ share}) \Bigr).
\end{align}
Therefore,
\begin{align}
\delta_{z,x}^{j,u}(0) &= \mathrm{CoVar}[R_{j,u}(0), R_{z,x}(0)] / M  \nonumber \\
&= jz \left( \sum_{k= \max \{-L, x-d_v+1\}}^{\min \{ L, u\}} \alpha^{L-|k|} \right) \nonumber \\
&\qquad \times \Bigl( P(d_u = j, d_x = z | \mathrm{share}) \nonumber \\
&\qquad \qquad - P(d_u = j, d_x = z | \mathrm{no \ share}) \Bigr). \label{ce_gsc_begin2}
\end{align}

When $j = d_c + 1$ and $z \leq d_c$, for $0 < x-u < d_v$, in a similar manner,
\begin{align}
\delta^{d_c+1,u}_{z,x}(0) &= \mathrm{CoVar}[V_u(0),R_{z,x}(0)] / M \nonumber \\
&= z \left( \sum_{m = z}^{d_c} \rho'_{m,x,\alpha} \binom{m-1}{z-1} \epsilon^{z} (1-\epsilon)^{m-z} \right) \nonumber \\
&\qquad \cdot \left( \sum_{m=z}^{d_c}\rho_{m,x,\alpha} \binom{m}{z}\epsilon^{z+1} (1-\epsilon)^{m-z}\right).
\end{align}
and otherwise,
\begin{align}
\delta^{d_c+1,u}_{z,x}(0) &= \mathrm{CoVar}[V_u(0),R_{z,x}(0)] / M = 0.\label{ce_gsc_end2}
\end{align}

\bibliographystyle{IEEEtran}
\bibliography{ref}

\begin{thebibliography}{1}
\providecommand{\url}[1]{#1}
\csname url@samestyle\endcsname
\providecommand{\newblock}{\relax}
\providecommand{\bibinfo}[2]{#2}
\providecommand{\BIBentrySTDinterwordspacing}{\spaceskip=0pt\relax}
\providecommand{\BIBentryALTinterwordstretchfactor}{4}
\providecommand{\BIBentryALTinterwordspacing}{\spaceskip=\fontdimen2\font plus
\BIBentryALTinterwordstretchfactor\fontdimen3\font minus
  \fontdimen4\font\relax}
\providecommand{\BIBforeignlanguage}[2]{{%
\expandafter\ifx\csname l@#1\endcsname\relax
\typeout{** WARNING: IEEEtran.bst: No hyphenation pattern has been}%
\typeout{** loaded for the language `#1'. Using the pattern for}%
\typeout{** the default language instead.}%
\else
\language=\csname l@#1\endcsname
\fi
#2}}
\providecommand{\BIBdecl}{\relax}
\BIBdecl

\bibitem{kudekar}
S.~{Kudekar}, T.~J. {Richardson}, and R.~L. {Urbanke}, ``Threshold saturation
  via spatial coupling: Why convolutional ldpc ensembles perform so well over
  the bec,'' \emph{IEEE Transactions on Information Theory}, vol.~57, no.~2,
  pp. 803--834, Feb 2011.

\bibitem{olmos_isit}
P.~M. Olmos and R.~Urbanke, ``Scaling behavior of convolutional ldpc ensembles
  over the bec,'' in \emph{2011 IEEE International Symposium on Information
  Theory Proceedings}, July 2011, pp. 1816--1820.

\bibitem{SFC}
Y.~NAKAHARA, S.~SAITO, and T.~MATSUSHIMA, ``Spatially ``mt. fuji'' coupled ldpc
  codes,'' \emph{IEICE Transactions on Fundamentals of Electronics,
  Communications and Computer Sciences}, vol. E100.A, no.~12, pp. 2594--2606,
  2017.

\bibitem{SFC_weight_distribution}
Y.~NAKAHARA and T.~MATSUSHIMA, ``A note on weight distributions of spatially
  ``mt. fuji'' coupled ldpc codes,'' \emph{IEICE Transactions on Fundamentals
  of Electronics, Communications and Computer Sciences}, vol. E101.A, no.~12,
  pp. 2194--2198, 2018.

\bibitem{SFC_EGE}
Y.~{Nakahara} and T.~{Matsushima}, ``Expected graph evolution for spatially
  ``mt. fuji'' coupled ldpc codes,'' in \emph{2018 International Symposium on
  Information Theory and Its Applications (ISITA)}, Oct 2018, p. 511.

\bibitem{olmos}
P.~M. Olmos and R.~L. Urbanke, ``A scaling law to predict the finite-length
  performance of spatially-coupled ldpc codes,'' \emph{IEEE Transactions on
  Information Theory}, vol.~61, no.~6, pp. 3164--3184, June 2015.

\bibitem{luby}
\BIBentryALTinterwordspacing
M.~G. Luby, M.~Mitzenmacher, M.~A. Shokrollahi, D.~A. Spielman, and V.~Stemann,
  ``Practical loss-resilient codes,'' in \emph{Proceedings of the Twenty-ninth
  Annual ACM Symposium on Theory of Computing}, ser. STOC '97.\hskip 1em plus
  0.5em minus 0.4em\relax New York, NY, USA: ACM, 1997, pp. 150--159. [Online].
  Available: \url{http://doi.acm.org/10.1145/258533.258573}
\BIBentrySTDinterwordspacing

\bibitem{amraoui}
A.~Amraoui, ``Asymptotic and finite-length optimization of ldpc codes,'' EPFL,
  Tech. Rep., 2006.

\end{thebibliography}

\end{document}